\documentclass[11pt,twoside]{article}

\usepackage{asp2006}
\usepackage{epsf}
\usepackage{psfig}
\usepackage{graphicx}
\usepackage{lscape}

\begin{document}
\title{ Extinction toward the Compact HII Regions G-0.02-0.07}

\author{Elisabeth Mills \altaffilmark{1}, Mark R. Morris \altaffilmark{1}, Cornelia C. Lang \altaffilmark{2}, \newline Angela Cotera \altaffilmark{3}, Hui Dong \altaffilmark{4}, Q. Daniel Wang \altaffilmark{4}, Susan Stolovy \altaffilmark{5} }  
\altaffiltext{1}{University of California, Los Angeles, Department of Physics and Astronomy, 430 Portola Plaza, Los Angeles, CA 90024. To contact: millsb@astro.ucla.edu} 
\altaffiltext{2}{The University of Iowa, 203 Van Allen Hall, Iowa City, IA 52242 } 
\altaffiltext{3}{SETI Institute, 515 N. Whisman Road, Mountain View, CA 94043 } 
\altaffiltext{4}{University of Massachusetts, 710 North Pleasant Street,Amherst, MA 01003 } 
\altaffiltext{5}{Spitzer Science Center, MS 220-6, Pasadena, CA 91125} 
\begin{abstract} 
The four HII regions in the Sgr A East complex: A, B, C, and D, represent evidence of recent massive star formation in the central ten parsecs.  Using Paschen $\alpha$ images taken with HST and 8.4 GHz VLA data, we construct an extinction map of A-D, and briefly discuss their morphology and location.
\end{abstract}

G-0.02-0.07 is a group of compact HII regions associated with M-0.02-0.07, the 50 km s$^{-1}$ cloud (Goss et al. 1985, hereafter G85). They appear to lie along a dense molecular ridge at its edge, which has apparently  been compressed by the Sgr A East supernova remnant (Serabyn et al. 1992, hereafter S92). Despite the suggestive arrangement of these sources along the periphery of Sgr A East, estimates of its age ($10^3-10^4$ years, Fryer et al. 2006, Mezger et al. 1989) suggest that the star formation predates the explosion. Studying the ongoing interaction of these structures may still provide clues to the origin of this star formation, but one is limited by difficulties in determining relative placements along the line of sight. To address this, we have used the extinction toward the HII regions to better constrain their location.

\begin{figure}[t!]
\includegraphics[width = 5.0in]{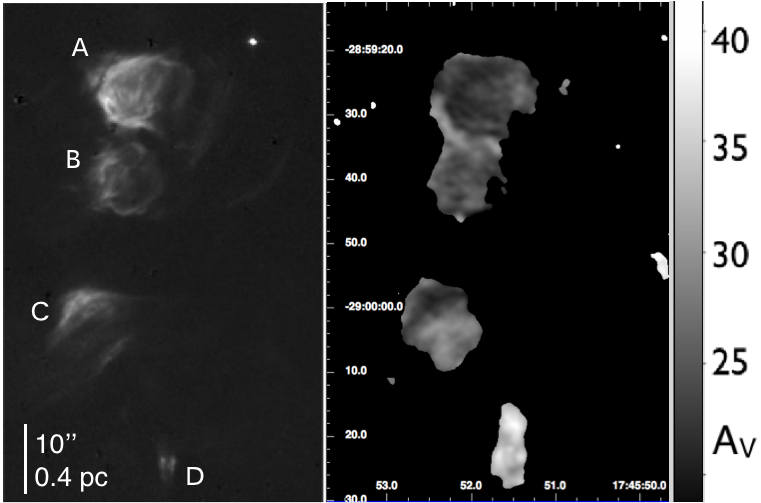}
\caption{Left: Paschen alpha map of the HII complex. Right: Extinction map, showing higher values of extinction for Source D. No information is available from these data on extinction values outside of these HII regions.}
\label{junk}
\end{figure}

An extinction map for these sources was constructed from the ratio of Paschen $\alpha$ and radio fluxes in the manner of Scoville et al. (2003). The radio image was assembled from unpublished 8.44 GHz data from the VLA archive (PI: Yusef-Zadeh). The data were taken between 1991-92 in all four array configurations, with a resulting beam size of $1.18"$x $0.66"$. The Pa $\alpha$ data were taken with NICMOS on HST, resolution $0.2"$, as part of a larger survey of the Galactic center (Wang et al. 2009). The images are constructed from differenced narrowband images at 1.87 and 1.9 $\mu$m to remove the stellar continuum emission. The radio data were also clipped at the 3 $\sigma$ level (0.9 mJy) before use in calculating the extinction. The extinction for A-C (A$_{V}\sim$ 25, see figure 1) is consistent with that found by Scoville et al. for Sgr A West, which appears to suffer no extinction from clouds in the central hundred parsecs. Source D (A$_{V}\sim35$) in contrast, appears more embedded.

The relatively low value of extinction toward A-C is consistent with a Galactic center distance, but suggests that these regions are located almost entirely in front of the 50 km s$^{-1}$ molecular cloud. We emphasize however that although A-C may currently be associated with little or no local attenuating material, we do not believe these sources lie in the foreground. S92 measure the velocities of A-D ranging from 43-46 km s$^{-1}$, and the CS peak of M-0.02-0.07 at 45 km s$^{-1}$, strongly suggesting the two are physically associated. Rather, we believe it is most likely that the HII regions are sufficiently evolved to have cleared much of their environment, consistent with observations by S92 of almost no overlap of CS emission from M-0.02-0.07 with the HII region positions. The additional ten magnitudes of extinction toward D suggest it is either embedded within or behind the compressed ridge at the edge of the cloud and Sgr A East. We believe that the former is the more likely of the two scenarios, as, like sources A-C, its velocity indicates it is associated with M-0.02-0.07.

\end{document}